\documentstyle[times,psfig]{l-aa}
\newcommand{\apj}[2]{ApJ #1, #2}
\newcommand{\apjs}[2]{ApJS #1, #2}

\newcommand{\aaa}[2]{A\&A #1, #2}

\newcommand{\jap}[2]{PASJ #1, #2}
\newcommand{\mon}[2]{MNRAS #1, #2}

\newcommand{\ergcm}[1]{10$^{#1}$ erg cm$^{-2}$ s$^{-1}$}
\newcommand{\ergs}[1]{10$^{#1}$ erg s$^{-1}$}
\newcommand{\hcm}[1]{10$^{#1}$ H cm$^{-2}$}
\newcommand{\rxj}{RX\,J0146.9$+$6121}
\newcommand{\xper}{X\,Persei}

\newcommand{\nh}{\hbox{N$_{\rm H}$}}
\newcommand{\et}{et al.}
\newcommand{\cts}{counts s$^{-1}$}

\begin{document}
 
\thesaurus{08;(08.05.2;  
               08.09.2;  
               08.14.1;  
               13.25.5)  
          }

\title{X-ray observations of the slowest known Be/X-ray pulsars \rxj\ 
       and \xper}
 
\author{F. Haberl\inst{1}, L. Angelini\inst{2,3}, C. Motch\inst{4} and 
        N.E. White\inst{2}}
 
\offprints{F. Haberl (fwh@mpe-garching.mpg.de)}
 
\institute{$^1$~Max-Planck-Institut f\"ur extraterrestrische Physik,
                Giessenbachstra{\ss}e, 85748 Garching, Germany \\
           $^2$~Laboratory for High Energy Astrophysics,
                NASA Goddard Space Flight Center, 
                Greenbelt, MD 20771, USA \\
           $^3$~University Space Research Association \\
           $^4$~Universite Louis Pasteur, Observatoire Astronomique de 
                Strasbourg, 11 rue de l'Universite, 67000 Strasbourg, France
             }
 
\date{Accepted 1 September 1997}
 
\maketitle\markboth{F. Haberl \et: X-ray observations of the slowest 
                    known Be/X-ray pulsars}
                   {F. Haberl \et: X-ray observations of the slowest 
                    known Be/X-ray pulsars}

\begin{abstract}
We present ASCA and ROSAT observations of the Be/X-ray binary pulsars
\rxj\ and \xper\ between 1990 and 1996. Measuring the neutron star spin
period and X-ray luminosity of \xper\ shows that the episode of low
X-ray luminosity and monotonic spin-down since 1978 continued 
to at least August 1995. ASCA and ROSAT HRI
observations of \rxj\ detected the pulsar with the longest known neutron 
star spin period at low luminosities after the X-ray outburst 
seen by EXOSAT in August 1984. The large spin period decrease seen 
between 1984 and 1993 has nearly stopped at a period of 1407.3 s in 
February 1996. We discuss that the X-ray outburst behaviour of \xper\ 
with respect to its Be star phase changes observed in the optical 
can be caused by an asymmetry in the matter distribution around the Be star.

\end{abstract}
 
 
\section{Introduction}

The low galactic latitude X-ray source \rxj\ was discovered by Motch \et\ 
(1991) from the ROSAT all-sky survey and identified with the Be star
LS\,I\,+61 235 (VES\,625). Further infrared and optical observations  
(Coe \et\ 1993, Motch \et\ 1997) confirmed the source as member of the 
Be/X-ray binaries, systems in which a compact object -- generally a 
neutron star -- accompanies a Be star in a wide eccentric orbit.
After ROSAT PSPC observations in February 1993 
revealed the pulse period of 1412 s (Hellier 
1994) it was clear that \rxj\ was responsible for the 25 min modulation 
in the X-ray flux seen in EXOSAT observations of the 8\,s 
pulsar 4U\,0142+61 (White \et\ 1987). \rxj\ is 
located only 24\arcmin\ away from this Uhuru source, well inside the 
collimator response of the EXOSAT ME instrument. 

The pulse period of 1412 s seen from \rxj\ exceeds the 837 s observed 
from \xper. These are far the longest known spin periods from a 
neutron star in an X-ray binary. 
\xper, a O9.5\,IIIe star (Slettebak 1982) has long been known as Be/X-ray 
binary. Recent optical observations revise the spectral classification 
to B0\,V and the distance to 700$\pm$300 pc (Lyubimkov \et\ 1997, Roche 
\et\ 1997).
After an X-ray outburst which peaked in 1975 and lasted probably 
more than 5 years the source shows an episode with modest X-ray luminosity of 
a few \ergs{34} and spin-down with $\dot P$/$P \sim 1.6$ 
10$^{-4}$ y$^{-1}$ since 1978 (for a review of the optical, IR and X-ray 
data see Roche \et\ 1993) at least until the ROSAT observation in 
August 1992 reported by Haberl (1994).

In this paper we present new X-ray observations of \rxj\ obtained with 
ASCA and ROSAT between 1990 and 1996 and ROSAT observations of \xper\ 
from February and August 1995. We detected \xper\ again at its low-level 
intensity, further extending the X-ray low-state. The X-ray 
behaviour of \rxj\ suggests large similarities of the two Be/X-ray 
binary systems.

\section{X-ray observations of \rxj}

\subsection{ASCA}

\begin{figure}
\psfig{figure=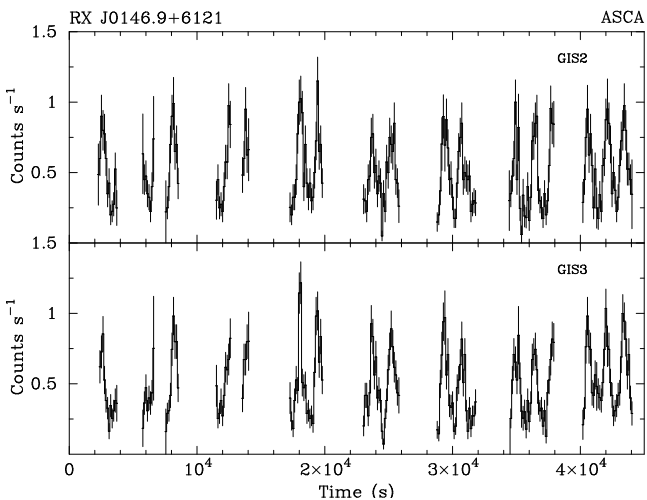,width=8.8cm,clip=,bbllx=25pt,bburx=520pt,bblly=20pt,bbury=395pt}
  \caption[]{0.8 $-$ 10 keV light curves of \rxj\ from the ASCA 
             GIS2 and GIS3 detectors starting on Sep. 18, 1994 at 
             22:18:54 UT. Each point is integrated over 120 s}
  \label{fig.rates0146asca}
\end{figure}

\begin{figure}
\psfig{figure=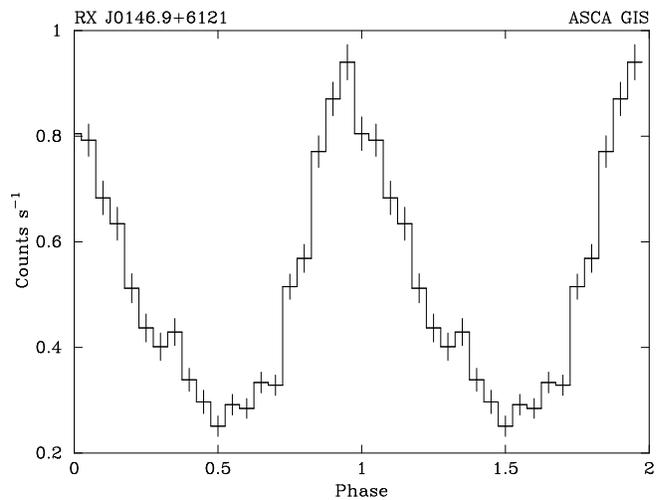,width=8.8cm,clip=,bbllx=25pt,bburx=520pt,bblly=25pt,bbury=405pt}
  \caption[]{Pulse profile of \rxj\ obtained in the 0.8 $-$ 10 keV 
             energy band by folding the GIS2 and GIS3 combined data with 
             a period of 1407.4 s. Two periods are plotted for clarity 
             versus arbitrary phase}
  \label{fig.puls0146asca}
\end{figure}

\rxj\ was observed in the 1 $-$ 10 keV band with ASCA 
(see Tanaka \et\ 1994) on September 18, 1994 
from 21:35 UT through 10:00 UT on the following day.
The prime target of the observation was the 8 s pulsar 4U\,0142+61 
(White et al. 1996).
The observation was arranged to include both sources in the 40\arcmin\ 
field of view (FOV) of the two gas imaging spectrometers, GIS2 and GIS3.
\rxj\ was located at off-axis angles of 
$\sim$17\arcmin\ and $\sim$13\arcmin\ for GIS2 and GIS3, respectively.
To avoid telemetry saturation the solid-state imaging spectrometer, SIS, 
did not include \rxj. 
The standard data selection filters were applied with a minimum elevation 
angle of 5 degrees, a cutoff rigidity of 6 GeV/c and rejection of events 
when the satellite was crossing the South Atlantic Anomaly. 
This yields a total exposure of 19839 s and 19855 s
for GIS2 and GIS3, respectively.
Source events were extracted within a 6\arcmin\ radius 
centered on the source positions and the background estimated from 
identically sized regions located in a source-free area of the GIS 
images. 
The vignetting corrected average count rate in the energy band 0.8 $-$ 10 keV
was $0.528 \pm 0.012$ and $0.510 \pm 0.009$ \cts\ for GIS2 and GIS3, 
respectively.
Figure~\ref{fig.rates0146asca} shows the light curves obtained by 
GIS2 and GIS3 where the $\sim$24 minute pulsations are directly visible. 
A pulse arrival time analysis was applied as described in White \et\ 
(1996) and a period of $1407.4 \pm 3.0$ s was derived.
The semi-amplitude of the modulation is $66 \pm 3 \%$ in the 0.8 $-$ 
10 keV energy range (Fig.~\ref{fig.puls0146asca}) and does not 
change significantly with energy.

\begin{figure}
\psfig{figure=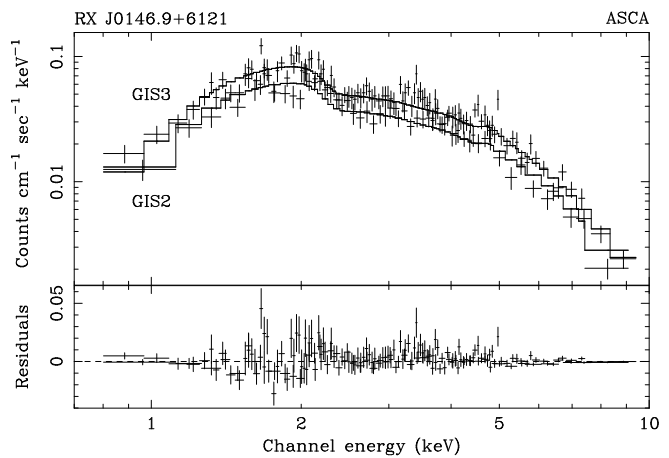,width=8.8cm,clip=,bbllx=25pt,bburx=520pt,bblly=25pt,bbury=365pt}
  \caption[]{ASCA GIS2 and GIS3 spectra (upper panel) plotted together 
             with the best fit power-law model as histogram. 
             Residuals are shown in the lower panel}
  \label{fig.spec0146asca}
\end{figure}

Energy spectra were extracted from the GIS2 and GIS3 data and re-binned to 
have at least 30 counts in each channel. 
Power-law, bremsstrahlung and blackbody single-component models including 
photoelectric absorption were simultaneously fit to the two spectra.
The uncertainties in the relative detector efficiencies were accounted for 
by including and fitting a relative normalization parameter.
Acceptable fits with a similar reduced $\chi^2$ of 1.1 were obtained with 
the power-law model, with a photon 
index of $1.46 \pm 0.08$ and absorption of 
$(1.00 \pm 0.15)$ \hcm{22}, and the bremsstrahlung model,
with a temperature of $24^{+12}_{-8}$ keV and a column 
density of $(0.91 \pm 0.1)$ \hcm{22}. The blackbody model gives a 
reduced $\chi^2$ of 1.4 with a column density of $0.0$.
From optical observations a E(B-V) of 1.09 is derived
(Motch \et\ 1997) yielding a column density of 6.3 $-$ 7.4 
\hcm{21} to \rxj. The power-law and bremsstrahlung model
suggest some system intrinsic absorption while the 
blackbody model can be rejected as unrealistic.
Figure~\ref{fig.spec0146asca} shows the GIS2 and GIS3 spectra and 
the residuals to the best fit power-law model.
No iron line was detected with an upper limit of 90 eV 
for the equivalent width of a narrow line at 6.4 keV.
The flux in the 0.5 $-$ 10 keV band estimated using the power-law model is 
3.9 \ergcm{-11}. 

Assuming that the spectral shape of \rxj\ between the ROSAT PSPC and ASCA 
observations did not change, we performed a simultaneous fit 
by allowing only the relative normalisation to vary. The absorbed 
power-law still gives an acceptable fit with reduced $\chi^2$ of 1.17. 
The absorption is 8.6 \hcm{21} and the photon index is 1.39. The 
intensity in the PSPC spectrum is only 3\% lower than the average 
derived for the GIS spectra, which is well within the errors.
An absorbed power-law with exponential high-energy cutoff was fit to
see if the spectral shape might be similar to that of 
\xper\ measured by BBXRT and ROSAT PSPC (Schlegel \et\ 1993, 
Haberl 1994). Such a model also fits the 0.1 $-$ 10 keV spectrum of 
\rxj\ with a reduced $\chi^2$ of 1.10. Formally the two more free 
parameters are not required but the model can not be excluded. The
best fit yields \nh\ = 6.0 \hcm{21}, photon index 0.69, cutoff energy 
$\sim$2 keV and folding energy 5.7, all very similar to the parameters 
found from \xper.

\subsection{ROSAT}

\rxj\ was observed by ROSAT in the 0.1 $-$ 2.4 keV energy band 
using the High Resolution Imager (HRI) 
as focal instrument between January 21 and February 28, 1996. 
The ROSAT mission is outlined by Tr\"umper (1983)
and the HRI is described by David \et\ (1993). 
The total net exposure of 38.7 ksec was distributed unevenly over the 
39 days with no observations from February 2$-$15 and 18$-$23. 

\begin{figure}
\psfig{figure=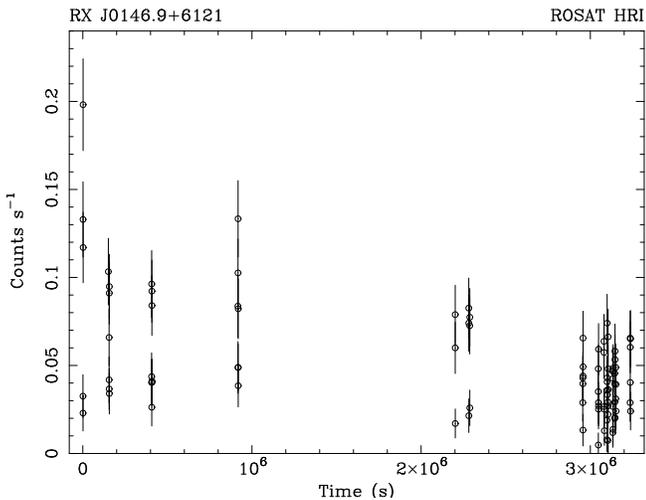,width=8.8cm,clip=,bbllx=25pt,bburx=520pt,bblly=20pt,bbury=395pt}
  \caption[]{ROSAT HRI light curve of \rxj\ obtained in Jan./Feb. 1996. 
             Each point represents an average over 300 s}
  \label{fig.rates0146hri}
\end{figure}

The source was detected in the ROSAT HRI observation with an average count 
rate of 0.046 $\pm$ 0.001 \cts\ and intensity variations between 0.005 and 
0.2 \cts, mainly caused by the X-ray pulsations. The 
light curve with a time resolution of 300 s is plotted in 
Fig.~\ref{fig.rates0146hri}.
Assuming a power-law spectrum as derived from the PSPC observation 
(Hellier 1994) the PSPC to HRI count rate conversion factor is 2.9.
This reveals an intensity decrease by about a factor of two between
the PSPC and HRI observation. This could be caused by an overall 
intensity decline or also by increased absorption.

\begin{figure}
\psfig{figure=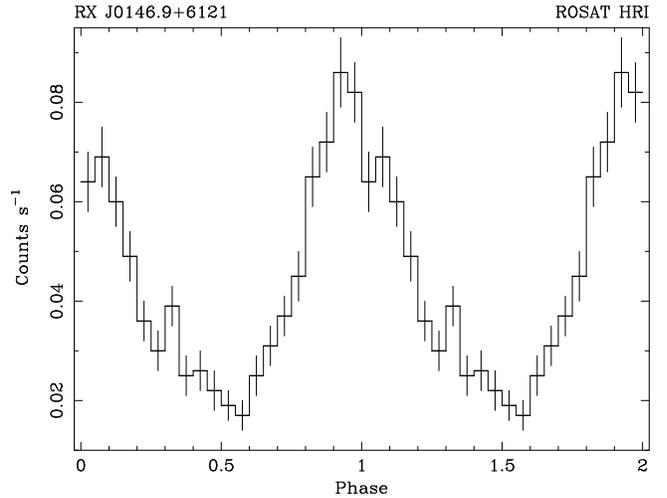,width=8.8cm,clip=,bbllx=25pt,bburx=520pt,bblly=20pt,bbury=395pt}
  \caption[]{Pulse profile of \rxj\ obtained from the ROSAT HRI 
             observation. The light curve is folded with 1407.28 s 
             and plotted as in Fig.~\ref{fig.puls0146asca}}
  \label{fig.puls0146hri}
\end{figure}

A pulse arrival time analysis was applied by splitting the HRI 
data into nine parts. Due to the long time span of the HRI observation 
the period can be determined accurately to 1407.28 $\pm$ 0.02 s. 
The folded light curve, shown in 
Fig.~\ref{fig.puls0146hri}, has a semi-amplitude modulation 
of 62 $\pm$ 7\% similar to the ROSAT PSPC and the ASCA observation.
The pulse profiles obtained from the HRI and the GIS observations are 
nearly identical and further illustrate the energy independence and
no changes with time. Small 
features on the decline part of the pulse peak are visible in the 
profiles of both instruments. The energy independence excludes 
absorption dips as origin and suggests some minor intensity peaks 
superimposed on the major pulse.

\subsection{History of pulse period and luminosity}

\begin{figure}
\psfig{figure=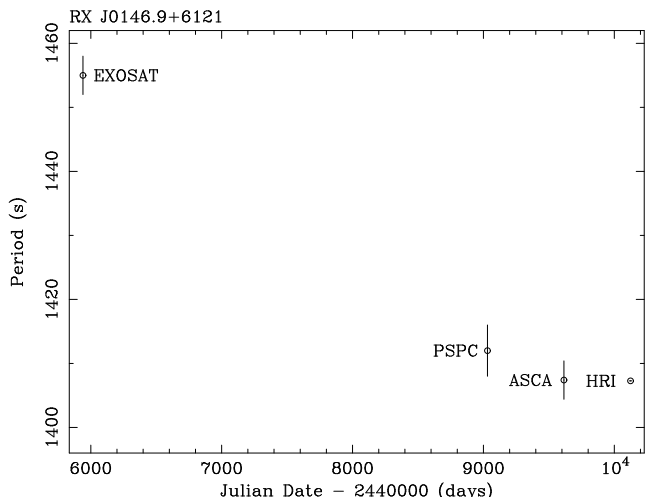,width=8.8cm,clip=,bbllx=25pt,bburx=520pt,bblly=20pt,bbury=395pt}
  \caption[]{Pulse period history of \rxj\ from August 1984 to February 1996}
  \label{fig.phist}
\end{figure}

The pulse period history of \rxj\ between the EXOSAT measurement in 
August 1984 and the ROSAT HRI point from February 1996 is summarized 
in Fig.~\ref{fig.phist}. 
The last two period measurements show that the large average period 
decrease of 5 s yr$^{-1}$ between 1984 and 1993 has slowed down 
considerably. 

The decrease in the spin-up rate of \rxj\ 
is accompanied by a fading in X-ray luminosity. Re-fitting the 
ROSAT PSPC spectrum with a power-law with the photon index fixed at the 
value derived from the ASCA spectrum yields a reduced $\chi^2$ of 1.2. 
The absorption is $(0.84 \pm 0.07)$ \hcm{22}, somewhat higher than 
the value derived by Hellier (1994), but consistent with the 
ASCA value. We therefore adopt the power-law index
of 1.46 for the spectrum of \rxj\ to calculate 0.5 $-$ 10.0 keV luminosities 
for comparison of the different observations assuming a
distance of 2.5 kpc. This is the distance of the open cluster in 
which \rxj\ is located (Tapia \et\ 1991). The luminosities 
corrected for absorption are 3.4 \ergs{34}, 4.0 \ergs{34} and 1.8 
\ergs{34} for the ROSAT PSPC, ASCA and ROSAT HRI observations, 
respectively. For the HRI observation no spectral change relative 
to the PSPC observation was assumed. 

During the ROSAT all-sky survey \rxj\ was 
scanned for about 3 days around August 1, 1990, 2.5 years before the ROSAT 
pointed observation. The average count rate during the total net 
exposure of 539 s was a factor of 1.8 below the count rate seen during 
the pointed observation (Motch \et\ 1997), i.e. at an intensity level 
comparable to the HRI observation. Again assuming the same spectrum as 
during the pointed PSPC observation the luminosity was about 1.9 \ergs{34}.

\section{ROSAT HRI observations of \xper}

\begin{figure}
\psfig{figure=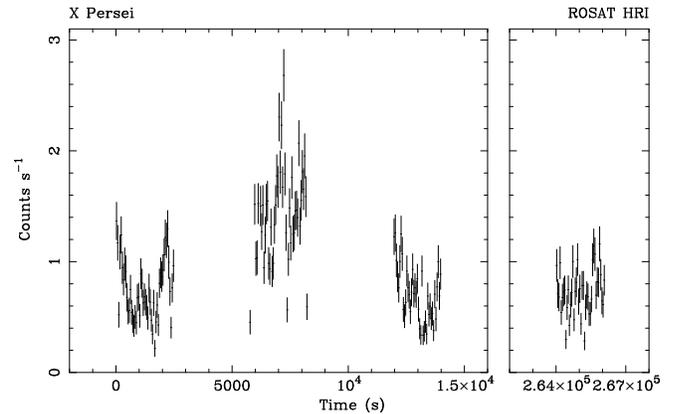,width=8.8cm,clip=,bbllx=25pt,bburx=510pt,bblly=55pt,bbury=360pt}
  \caption[]{HRI light curve of \xper\ from August 16$-$20, 1995 with a time 
             binning of 50 s. Note the gap of nearly 3 days between the last two 
             observation intervals}
  \label{fig.rates_xper}
\end{figure}

\xper\ was re-observed by ROSAT using the HRI as focal plane detector on
February 28, 1995 between 05:31 and 10:14 UT for a net exposure time of
7743 s and from August 16, 1995 23:32 UT to August 20, 1995 01:28 UT 
for 9829 s. The average count rates were 0.84 $\pm$ 0.01 \cts\ 
and 0.85 $\pm$ 0.01 \cts, respectively. The light curve of the August observation is shown
in Fig.~\ref{fig.rates_xper}. The temporal behaviour is similar to the
PSPC observation in 1992 (Haberl 1994) with the pulse modulation on top
of factor $\sim$2 variations with time scales of around 2 hours.
Assuming the power-law spectrum best representing the PSPC spectrum
(photon index 0.64 and \nh\ = 1.4 \hcm{21}), an average HRI count rate
of 1.0 \cts\ is expected. The lower observed HRI count rate may be
caused by higher absorption as it was e.g. observed during the BBXRT
observation (\nh\ = 2.5 \hcm{21}, Schlegel \et\ 1993) or by a lower
intrinsic X-ray intensity. Thus, unless the X-ray spectrum has 
changed dramatically, the X-ray luminosity is in the range 3$-$4
\ergs{34} (2$-$10 keV using a distance of 1300 pc for comparison with 
previous measurements), at the
same level \xper\ was always detected after its outburst in 1975 
(Roche \et\ 1993, Haberl 1994).

A timing analysis of the \xper\ HRI observations 
performed in the same way as for \rxj\ yields a pulse period of 837.2 
$\pm$ 0.1 s for the August 1995 observation. The observation 
from February 1995 was too short and did not cover a long enough base 
line to constrain the period. The new HRI measurement further 
continues the spin-down episode of \xper\ with $\dot P$/$P \sim 1.5$ 
10$^{-4}$ y$^{-1}$.

\begin{figure}
\psfig{figure=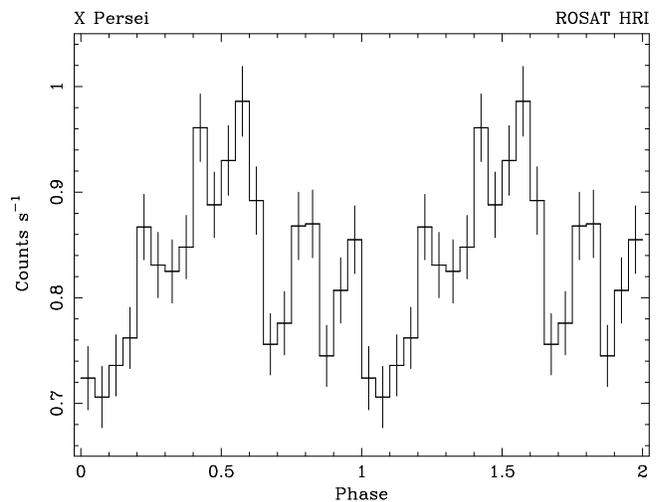,width=8.8cm,clip=,bbllx=25pt,bburx=520pt,bblly=20pt,bbury=395pt}
  \caption[]{Pulse profile of \xper\ obtained by combining the two HRI 
             observations half a year a part and assuming a constant 
             $\dot P$ of 2.4 10$^{-9}$ s s$^{-1}$}
  \label{fig.puls_xper}
\end{figure}

To improve the statistics of the pulse profile the two HRI observations 
were combined by assuming a constant $\dot P$ value of 
2.4 10$^{-9}$ s s$^{-1}$ as
it was observed between the PSPC and the second HRI observation. 
The profile is shown in Fig.~\ref{fig.puls_xper}. Although the 
statistical quality is still low the profile is consistent with that 
obtained from the PSPC observation.

\section{Discussion}

X-ray observations of the 1400 s pulsar \rxj\ from ASCA and
ROSAT between 1990 and 1996 show the source to be at a low 
luminosity around 2 \ergs{34} (0.5 $-$ 10 keV, 2.5 kpc). 
This is a factor of 20 fainter than the outburst luminosity seen 
by EXOSAT in August 1984. A large decrease in pulse period 
from 1455 s to 1412 s occured between 1984 and 1993 (Hellier 1994).
During the ASCA and ROSAT observations, September 1994 and 
February 1996, the period was measured to be 1407.4 s and 1407.3 s, 
respectively. The last
X-ray observations suggest that the spin period of the neutron star in
\rxj\ has reached a relatively constant value of 1407 s for an X-ray
luminosity of a few \ergs{34}. This period is unlikely the Kepler
period at the inner accretion disk which is bounded by the neutron stars
magnetosphere to be in accretion equilibrium, as it would imply a
magnetic field strength of 8 10$^{13}$ G for a standard 1.4 M$_\odot$
neutron star with 10 km radius. This is much higher than magnetic field
strength values derived from cyclotron features seen in the X-ray
spectra of X-ray pulsars which only in the case of A0535+26 may reach
10$^{13}$ G (Grove \et\ 1995). For more typical values of B around
10$^{12}$ G the equilibrium period for a luminosity of 2 \ergs{34} is
expected to be around 40 s, far from the observed period. 

The monotonic spin-down episode
in \xper\ indicates that any accretion disk is transitory, only
forming during outbursts. Standard accretion disk theory then predicts 
(e.g. Frank \et\ 1992) spin-up as was observed from \xper\ during the 
outburst.
The long spin period indicates that the
episodes of spin-down dominate the spin period changes and the increased
accretion via a disk during rare outbursts does not bring 
the spin into equilibrium. If a similar scenario is valid for
\rxj, the low $-$ still slightly decreasing $-$ X-ray luminosity and the
much reduced spin-up observed in the last years suggests that the
accretion disk has dissipated and that it may also enter an episode of
low X-ray luminosity and spin-down.

Evolutionary considerations suggest that Be\,$-$\,neutron star 
binary systems are formed from close binaries of massive stars (van den 
Heuvel 1983).
\xper\ and \rxj\ are probably systems with long orbital periods
of the order of several hundreds of days. A 580 day period claimed 
for \xper\ (Hutchings \et\ 1974) could however not be confirmed.
One such long orbital period system is PSR\,B1259$-$63, a 47\,ms pulsar 
in a 1237 day orbit around a Be star. If the neutron stars in \xper\ and 
\rxj\ were born with similar short spin period and their present long
spin periods are due to spin-down they must be older systems.

The ASCA spectrum of \rxj\ is the first measured broad band spectrum 
(0.8 $-$ 10 keV) without confusion by the close pulsar 4U\,0142+61.
It is consistent with a power-law with photon index 1.5 which is 
relatively steep for X-ray pulsars (typical 0.8 $-$ 1.5, Nagase 1989).
\xper, the other long period pulsar, also shows a steep spectrum with a 
photon index of 0.8 and an exponential cutoff above 2.2 keV. This cutoff-model 
was suggested by BBXRT data (Schlegel \et\ 1993) and the break confirmed 
by the flatter ROSAT PSPC spectrum in the 0.1 $-$ 2.4 keV range (Haberl 
1994). Combining the ROSAT PSPC and the ASCA GIS spectra of \rxj\ in a 
simultaneous model-fit shows that the spectrum can also be represented 
by a power-law with exponential 
cutoff as in the case of \xper. The power-law index, cutoff and folding energy 
(0.69, 2 keV and 5.7 keV, respectively) are consistent within the errors 
with those found from \xper. 

It has been suggested that the cutoff 
seen in the power-law spectra of X-ray pulsars is related to the 
magnetic field strength of the neutron star. Observed cyclotron 
line energies E$_0$ and cutoff energies E$_c$ are correlated 
with E$_0$ $\sim$ 2 E$_c$ (Makishima \& Mihara 1992, see also White \et\ 
1993). This would suggest field strengths of a few 10$^{11}$ G for 
\xper\ and \rxj\ (if there is a break in the spectrum), one order of 
magnitude lower than for typical X-ray pulsars which show cutoff energies 
of 10 $-$ 20 keV. However, no cyclotron features have been directly 
seen in the spectra of \xper\ and \rxj. At these relatively low magnetic 
fields strengths the distortion of the spectrum from the neutron star 
surface will be much reduced, and the spectrum should resemble a 
blackbody (White \et\ 1996). A luminosity dependence of the cutoff 
energy seen from EXO\,2030$+$375 also indicates that the cutoff energy may 
not provide a reliable measure of the surface magnetic field strength
(Reynolds \et 1993).

\rxj\ and \xper\ show persistent low-level X-ray flux
after their outbursts with little variations.
Both sources have similar luminosities in 
the 2 $-$ 10 keV band (\rxj\ $\sim$ 1.2$-$2.9 \ergs{34}, 
\xper\ $\sim$ 0.7$-$2.5 \ergs{34}).
For \rxj\ a distance of 2.5 kpc is used, the distance to the 
open cluster \rxj\ is probably associated with (Tapia \et\ 1991).
Motch \et\ (1997) give 2.9 kpc but Reig \et\ (1997) derive 
a spectral type and luminosity class of B1V, more consistent with 2.5 kpc.
We use a revised distance estimate for \xper\ of 700 pc 
(Lyubimkov \et\ 1997, Roche \et\ 1997).
This relatively constant X-ray luminosity 
needs to be explained in the framework of stellar wind accretion.

\xper\ is well observed in the optical and large changes in the 
morphology of the circum-stellar envelope are indicated.
The V-band light curve shows 
extended faint states and spectroscopic observations revealed a phase 
transition of the Be star to a normal B star in the last extended low 
state 1989 to 1993 (Roche \et\ 1993, 1997) which indicates the 
disappearance of the equatorial low-velocity and high-density outflow 
around the Be star. 
The X-ray outburst of \xper\ in 1975 
happened during an optical extended low-state, suggesting that the 
dissipating circum-stellar matter crossed the neutron star and leading 
to enhanced accretion. In contrast 
no similar X-ray outburst was observed
during the last extended optical low-state. 
One possibility may be that the 
orbital plane is inclined against the equatorial plane of the Be star 
with the neutron star well outside the equatorial dense outflow 
during the last 
phase transition. However UV observations of Be stars indicate a 
low-density wind at higher latitudes and the derived low 
mass loss rates of the order of 10$^{-9}$ to 10$^{-8}$ M$_\odot$ y$^{-1}$ 
and high outflow velocities up to 2000 km s$^{-1}$ can not account for 
the observed X-ray luminosity of \ergs{34} (see e.g. Waters \et\ 
1988). If the neutron star moves in the equatorial plane a radially 
outward moving dense wind region can only miss the neutron star 
if it is highly non-axisymmetric. Long term variations in the ratio of 
the intensity of the violet and red peak of the Balmer emission lines is 
commonly observed from Be stars, consistent with such an asymmetry in the 
matter distribution around the star (e.g. $\gamma$ Cas, Telting \& Kaper 1994).
In particular the optical counterpart of \rxj\ shows that either the 
violet or the red part of the line can be in absorption (Motch \et\ 
1997, Reig \et\ 1997) indicating that the dense wind outflow responsible for 
Balmer emission is sometimes located only on one side of the star.
The relatively constant X-ray luminosity then suggests that even in the 
case of a "complete loss'' of the dense equatorial stellar envelope as 
indicated by Balmer lines in absorption, a basic high-density 
low-velocity outflow is permanently present to power the X-ray source.
This picture also suggests that the frequency of X-ray outbursts in wide 
Be/X-ray binaries is lower than that of Be\,$-$\,B star phase transitions.
To prove this, further monitoring of \rxj, \xper\ and other 
low-luminosity Be/X-ray binaries like those discovered by ROSAT (Motch 
\et\ 1997) is required.
\acknowledgements 
The ROSAT project is supported by the German Bundesministerium f\"ur
Bildung, Wissenschaft, Forschung und Technologie (BMBF/DARA) and the
Max-Planck-Gesellschaft. CM acknowledges support from a CNRS-MPG 
cooperation contract. NEW thanks the US ASCA Guest Observer facility 
staff for their assistance in enabling the observation of two
pulsars.

\end{document}